\documentclass[aps,prc,reprint,superscriptaddress,amsmath,amssymb]{revtex4-2}
\usepackage{graphicx}
\usepackage{hyperref}
\usepackage{orcidlink} 
\usepackage{amsmath}
\usepackage{float}
\hypersetup{
  colorlinks=true,
  linkcolor=blue,
  citecolor=blue,
  urlcolor=blue,
}

\DeclareSymbolFont{lettersA}{U}{txmia}{m}{it}
\DeclareMathSymbol{\varv}{\mathord}{lettersA}{51}

\newcommand{\nalpha}{\texorpdfstring{\ensuremath{\mathrm{n}+\alpha}}{n+α}}
\newcommand{\noxygen}{\texorpdfstring{\ensuremath{\mathrm{n}+{}^{16}\mathrm{O}}}{n+¹⁶O}}
\newcommand{\pwave}{\texorpdfstring{\ensuremath{P_{3/2}}}{P3/2}}
\newcommand{\dwave}{\texorpdfstring{\ensuremath{D_{3/2}}}{D3/2}}

\begin{document}
\title{Neural Network Potential Reconstruction from Phase Shifts in Resonant \noxygen{} Scattering}
\author{Lucas~A.~Souza\,\orcidlink{0000-0002-8711-956X}}
\email{lasouza@if.usp.br}
\affiliation{Independent Researcher, Minas Gerais, Brazil}

\author{Edson~Otoniel\,\orcidlink{0000-0001-9929-5977}}
\email{edson.otoniel@ufca.edu.br}
\affiliation{Universidade Federal do Cariri, Instituto de Forma\c c\~ao de Educadores,
Brejo Santo, CE, Brazil}

\begin{abstract} 
We reconstruct a local, energy-independent effective potential for the narrow \dwave{} resonance of \noxygen{} from elastic phase shifts, using a Gaussian-enveloped neural parametrization coupled to a differentiable variable-phase calculation and checked by independent radial Schr\"odinger solvers. For the first resonance (below \(2.5\) MeV LAB), the reconstruction gives a \(90^\circ\) crossing at \(1.001\) MeV LAB and a derivative width of \(85\) keV LAB. The published Dubovichenko potential gives similar values with its original kinetic coefficient. Four reconstructions with adjustable range place the crossing within \(5\) keV of one another while their central potentials span \(47\) MeV. This spread illustrates the radial ambiguity of the inversion for the measured \dwave{} phases. The broad \pwave{} resonance of \nalpha{} serves as a benchmark for the same method.
\end{abstract}
\maketitle

\section{Introduction}

Elastic-scattering phase shifts constrain effective nuclear
interactions and resonance properties. The variable-phase
formulation relates an interaction potential directly to its
phase shifts~\cite{Babikov1967,Calogero1967}. For nucleon--alpha
scattering, Alexander \textit{et al.} inverted phase shifts with
a modified Newton--Sabatier method and obtained radial forms
distinct from the phenomenological potentials that generated some
of their input phases~\cite{Alexander1996}. Energy-dependent
nucleon--alpha potentials have also been obtained by \(S\)-matrix
inversion~\cite{Cooper1996}. These inversions
motivate asking which radial features can be inferred from
elastic phases in a finite energy interval and a single partial
wave.

Machine learning is now widely applied across nuclear
physics~\cite{Boehnlein2022}, including forward scattering and
cross-section problems~\cite{Lei2026,LeiOptical2026,Xing2024}.
Computational approaches to the scattering
inverse problem~\cite{ChadanSabatier1977} include
Balassa's Volterra-series and neural-network reconstructions for
neutron--alpha scattering~\cite{Balassa2022,Balassa2023,Balassa2024},
and the reference-potential methods of Sastri and collaborators
for resonant channels~\cite{Sastri2024,Awasthi2025}. Most closely
related to the present work, Awasthi \textit{et al.} reconstructed
the \pwave{} \nalpha{} interaction with a Gaussian-enveloped neural
potential and differentiable integration of the variable-phase
equation~\cite{Awasthi2026}. We adopt that construction, use
\nalpha{} as a benchmark, and extend the analysis to the narrow
\dwave{} resonance of \noxygen{}.

The \dwave{} resonance near 1 MeV in \noxygen{}, established and
interpreted microscopically by George and Philpott in a continuum
shell model~\cite{GeorgePhilpott1974,GeorgePhilpott1975}, is the
structure we reconstruct here directly from the elastic phases,
using a local effective potential. Local potentials for
\noxygen{} have been extracted before by inverting the exchange
nonlocality at higher energies~\cite{Rawitscher1994,Lukaszek1996}, and
the phase-function method has been applied to the same
system~\cite{Sharma2023}; our target is instead the narrow \dwave{}
resonance. Its narrowness makes
the calculated phase hypersensitive to the location of its rapid
rise, placing demands on the branch treatment and on the Hamiltonian
that a broad resonance does not impose.

In this work we reconstruct local, energy-independent effective
potentials from elastic phase shifts by optimizing a neural
potential through a differentiable variable-phase calculation, with
the resulting phases checked by independent radial Schr\"odinger
solvers. The principal application is the narrow \dwave{} resonance
of \noxygen{}, with the data restricted to the first resonance: we
recover its observables, examine the stability of the reconstruction
across physically informed starts, and find that the elastic phases
locate the phase crossing more tightly than they constrain the central
depth among the solutions examined. We also evaluate the sensitivity of
the crossing to the kinetic coefficient at fixed potential. For the
broad-resonance benchmark we first apply the same method to the
\pwave{} channel of \nalpha{}~\cite{Awasthi2026} and compare with
established interactions~\cite{Satchler1968,KKNN1979}.

\section{Theoretical and Computational Framework}
\label{sec:framework}

\subsection{Scattering formalism}

We reconstruct a real, local, energy-independent nuclear potential
\(V_{\ell j}(r)\) for elastic neutron scattering from a spin-zero
target in the \(P_{3/2}\) channel of \nalpha{} and the \(D_{3/2}\)
channel of \noxygen{}. For neutron mass \(m_n\) and target mass
\(M_T\), the reduced mass is \(\mu=m_nM_T/(m_n+M_T)\). With the
target initially at rest, \(E_{\mathrm{cm}}=M_T E_{\mathrm{lab}}/(m_n+M_T)\)
and \(k=\sqrt{2\mu E_{\mathrm{cm}}}/\hbar\). The reduced radial
wave function satisfies
\begin{equation}
\label{eq:radial-schrodinger}
u_{\ell j}''+
\left[k^2-\frac{\ell(\ell+1)}{r^2}
-\frac{2\mu}{\hbar^2}V_{\ell j}(r)\right]u_{\ell j}=0,
\end{equation}
with the centrifugal term explicit. We use
\(\hbar c=197.3269804\) MeV fm. The helium calculation uses rest
energies of \(939.56542052\) and \(3727.379378\) MeV for the neutron
and \(\alpha\) particle. The oxygen reconstruction uses
\(m_n=1.00866491606\) u, the atomic target mass
\(M_T=15.99491461957\) u, and \(1\,\mathrm{u}\,c^2=931.49410372\) MeV.

\begin{figure*}[t]
\centering
\includegraphics[width=\textwidth]{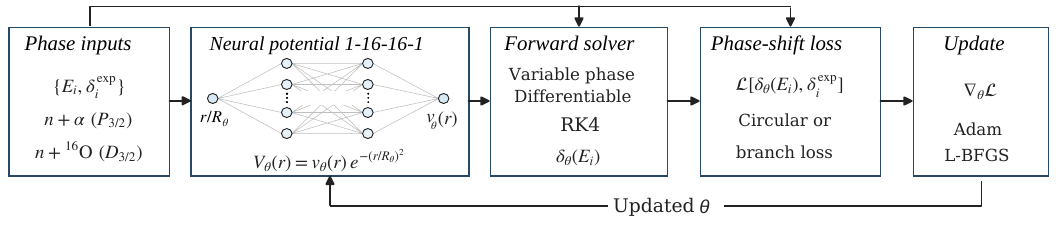}
\caption{Neural inverse-scattering workflow. At the input energies,
differentiable RK4 maps the Gaussian-enveloped potential to phase
shifts. The phase loss drives the parameter update. The MLP supplies
the amplitude for helium and the output \(N_\theta(x)\) used to
construct the oxygen amplitude in Eq.~\eqref{eq:oxygen-potential}.
Independent Numerov (helium) and Schr\"odinger/DOP853 (oxygen)
calculations contribute to certification.}
\label{fig:inverse_workflow}
\end{figure*}

The regular solution behaves as \(r^{\ell+1}\) at the origin and
has the asymptotic form
\begin{equation}
u_{\ell j}(r)\propto
\hat j_\ell(kr)\cos\delta_{\ell j}
-\hat n_\ell(kr)\sin\delta_{\ell j},
\end{equation}
where \(\hat j_\ell(x)=xj_\ell(x)\) and \(\hat n_\ell(x)=xy_\ell(x)\)
are the Riccati--Bessel and Riccati--Neumann functions built from
the spherical functions \(j_\ell\) and \(y_\ell\). The variable-phase equation~\cite{Babikov1967,Calogero1967} gives
the phase directly from the potential. Suppressing the channel
indices on the running phase, it reads
\begin{equation}
\label{eq:variable-phase}
\frac{d\delta(r;k)}{dr}=-\frac{2\mu V_{\ell j}(r)}{\hbar^2 k}
\bigl[\hat j_\ell(kr)\cos\delta
     -\hat n_\ell(kr)\sin\delta\bigr]^2 .
\end{equation}
The regular condition is \(\delta(0;k)=0\). The numerical
propagator starts with zero phase at a small positive radius
and approximates the asymptotic phase by its value at the outer
node. The elastic element \(S_{\ell j}=e^{2i\delta_{\ell j}}\)
is invariant under \(\delta_{\ell j}\mapsto\delta_{\ell j}+m\pi\)
for integer \(m\). Differentiating the propagator supplies the
phase sensitivities for the inverse fit (Fig.~\ref{fig:inverse_workflow}).

\subsection{Neural representation of the interaction potential}

The nuclear potential in each channel is
\begin{equation}
\label{eq:neural-potential}
V_\theta(r)=\varv_\theta(r)\exp[-(r/R_\theta)^2],
\end{equation}
where the Gaussian envelope sets the decay scale \(R_\theta\)
and the neural amplitude \(\varv_\theta(r)\) can vary with radius.
A fully connected \(1\)--\(16\)--\(16\)--\(1\) multilayer
perceptron (MLP) takes the dimensionless input \(x=r/R_\theta\).
Both hidden layers use \(\tanh\), and the output is linear and
interpreted directly in MeV.

Only the way the MLP defines the amplitude differs between the two channels.
For helium the MLP output is the amplitude itself, \(\varv_\theta(r)\),
so the network is free at every radius. For oxygen we instead
constrain the origin,
\begin{equation}
\label{eq:oxygen-potential}
V_\theta(r)=\bigl[V_c+x^2N_\theta(x)\bigr]e^{-x^2},
\qquad x=r/R_\theta ,
\end{equation}
where the MLP output \(N_\theta(x)\) enters only through the \(x^2\)
correction, so that \(V_\theta(0)=V_c\) and \(V_\theta'(0)=0\). This
smooth, regular interior stabilizes the more delicate narrow-resonance
fit; the amplitude is then \(\varv_\theta(r)=V_c+x^2N_\theta(x)\).

For helium, a preliminary phase fit supplies the Gaussian
depth and range. Zero output weights and a bias equal to
the fitted depth give a constant initial amplitude. For oxygen,
three distinct depth--range pairs from a certified resonant
region initialize \(V_c\) and \(R_\theta\). A zero output
layer gives a vanishing radial correction during the initial
adjustment of depth and range. The neural stage in both systems
optimizes all network weights and biases together with a logarithmic
range coordinate and, for oxygen, \(V_c\). The envelope restricts
the family of reconstructed potentials while leaving their radial
shape adjustable. Fitting the phases does not establish uniqueness.

\subsection{Inverse optimization}

In both systems the fit minimizes the mean-squared mismatch between
the calculated and reference phase shifts at the input energies; the
systems differ only in how that mismatch is measured. Writing
\(\Delta_i=\delta_i^{\mathrm{calc}}-\delta_i^{\mathrm{ref}}\) in radians,
helium uses the circular loss
\begin{equation}
\label{eq:circular-loss}
\mathcal L_{\mathrm{circ}}=\frac{4}{N}\sum_i\sin^2\Delta_i ,
\end{equation}
the mean squared distance between the elastic \(S\)-matrix elements,
which is invariant under \(\delta\mapsto\delta+\pi\). For the narrow
oxygen resonance the phase climbs rapidly through \(\pi/2\), so there
we fit the continuous (unwrapped) phase branch,
\begin{equation}
\label{eq:branch-loss}
\mathcal L_{\mathrm{branch}}=\frac{1}{N}\sum_i
\bigl(\widetilde\delta_i^{\mathrm{calc}}
-\widetilde\delta_i^{\mathrm{ref}}\bigr)^2 .
\end{equation}
The calculated branch is obtained by unwrapping twice the phase in
increasing energy (correcting jumps larger than \(\pi\) in
magnitude), dividing by two, and adding a common multiple
of \(\pi\) to place the first value nearest \(-\pi\). The
Fowler--Cohn input phases are shifted by \(-\pi\) and Hickey's
are used as published. The branch integers are recomputed at each
trial and held fixed in the local derivative. The two energy grids
contain the observation energies and have nominal spacings of 2.5
and 5 keV LAB, starting at 5 and 10 keV LAB, respectively. Agreement
between their branches, proximity to the low-energy anchor, and
bounds on phase increments and changes between trials guard
against branch errors near narrow resonances.

Each training loss averages the source means equally. For helium
the sources are 22 Morgan--Walter and 20 Hoop--Barschall
observations and 22 phases calculated with the Satchler optical
model~\cite{Morgan1968,Hoop1966,Satchler1968}. Oxygen uses six
Fowler--Cohn and seven Hickey observations~\cite{Fowler1958,Hickey1974}.
Evaluation pools points within the reported sample using
\begin{equation}
\label{eq:circular-rmse}
\mathrm{RMSE}_{\mathrm{circ}}
=\left[\frac{1}{N}\sum_i
\bigl(\tfrac12\operatorname{Arg}e^{2i\Delta_i}\bigr)^2
\right]^{1/2}.
\end{equation}
The RMSE is reported in degrees for the 42 helium experimental
points and all 13 oxygen observations, using Numerov and DOP853
phases, respectively. Source-specific RMSEs pool only the
observations of the indicated source.

The training forward calculation integrates
Eq.~\eqref{eq:variable-phase} with RK4 in float64 on a CPU.
Derivatives of the discrete RK4 steps are composed with automatic
derivatives of the potential. For oxygen, batched propagation
assigns the branch on the dense energy mesh and scalar propagation
supplies sensitivities at the observation energies.

Both neural fits use Adam at learning rate \(10^{-3}\) followed
by L-BFGS~\cite{LiuNocedal1989} with Armijo backtracking.
Helium uses 2000 Adam updates and up to 300 L-BFGS trial
evaluations. The selected state minimizes the training loss
among certified initial, periodic, and terminal states.

For oxygen, a coarse Gaussian depth--range scan and one local
refinement locate a region with a certified resonance near
1 MeV LAB. The three distinct starts are the eligible certified
points with \(E_{90}=1.00\pm0.01\) MeV LAB, after excluding
two previously used reference points. This physically informed
initialization targets the resonant region. Training minimizes
the phase loss alone. The initial depth and range adjustment
uses up to 50 Gauss--Newton proposals with the network fixed.
Joint optimization then uses up to 1000 Adam proposals and
150 L-BFGS evaluations. Each proposal limits changes in
\(V_c\) and \(R_\theta\) to 0.5 MeV and 0.005 fm. In both
systems, numerically invalid trials are rejected and the step
is reduced by halving. Rejections count against the oxygen
budgets. For each oxygen start, the lowest-loss certified
state among the depth--range, post-Adam, and terminal states
is recertified. The campaign also requires an in-band
\(E_{90}\), a circular RMSE of at most \(6^\circ\), and
a nonzero radial correction.

Certification checks radial step halving and extension of the
integration domain. The helium RK4 grid spans 0.01--18 fm with
nominal spacing 0.00125 fm and is extended to 22 fm. The oxygen
grid spans 0.001--15 fm with nominal spacing 0.0025 fm and is
extended to 20 fm. Independent radial solutions use Numerov for
helium and DOP853 for oxygen, matched to free Riccati functions.
Phase agreement is tested to \(0.01^\circ\), with an additional
\(|\Delta S|<10^{-4}\) check for oxygen. Helium checks the
stability of the raw phase under radial refinement. Oxygen
compares the dense branch against scalar RK4, radial refinement,
and sampled DOP853. Gradient controls use radial refinement and
centered finite differences, including a branch-checked log-range
audit for oxygen. Finite-box radial spectra check negative-energy
levels in both channels. Oxygen certification also requires
the absence of negative \(D\)-wave levels on the tested meshes
and verifies the value and slope at the origin.

For helium, the resonance observables are the peak energy of
the partial cross section, proportional to \(\sin^2\delta/k^2\),
and its FWHM, the separation of the two half-height energies
without background subtraction. They are reported in MeV CM.
The oxygen observable \(E_{90}\) is the upward
\(-\pi/2\) crossing of the chosen branch, refined with DOP853
in 0.5--2.5 MeV LAB. The derivative width is
\begin{equation}
\Gamma_\delta=
\left.\frac{2}{d\widetilde\delta/dE_{\mathrm{lab}}}\right|_{E_{90}} .
\end{equation}
The derivative is in radians per MeV LAB and the width is
reported in keV LAB. The energy dependence of \(k^{-2}\) can
shift the cross-section peak away from \(E_{90}\). The
derivative width measures the phase slope at the crossing,
whereas the FWHM measures the span of the cross-section peak.

\raggedbottom
\section{Results}

\subsection{Broad-resonance benchmark: \nalpha{} \pwave{}}
We first apply the method to the broad \pwave{} resonance of
\nalpha{}, where the closest neural precedent exists~\cite{Awasthi2026}
and established interactions are available for comparison.
The \(P_{3/2}\) fit uses 22 experimental phases from Morgan and
Walter~\cite{Morgan1968}, 20 from Hoop and Barschall~\cite{Hoop1966},
and 22 phases calculated by the Satchler optical model~\cite{Satchler1968}.
The calculated phases supply model input. The circular RMSE in Table~\ref{tab:nalpha_comparison} uses only the
42 measured phases. These points also enter the fit, so the metric
measures in-sample agreement.

\begin{figure}[H]
\centering
\includegraphics[width=\columnwidth]{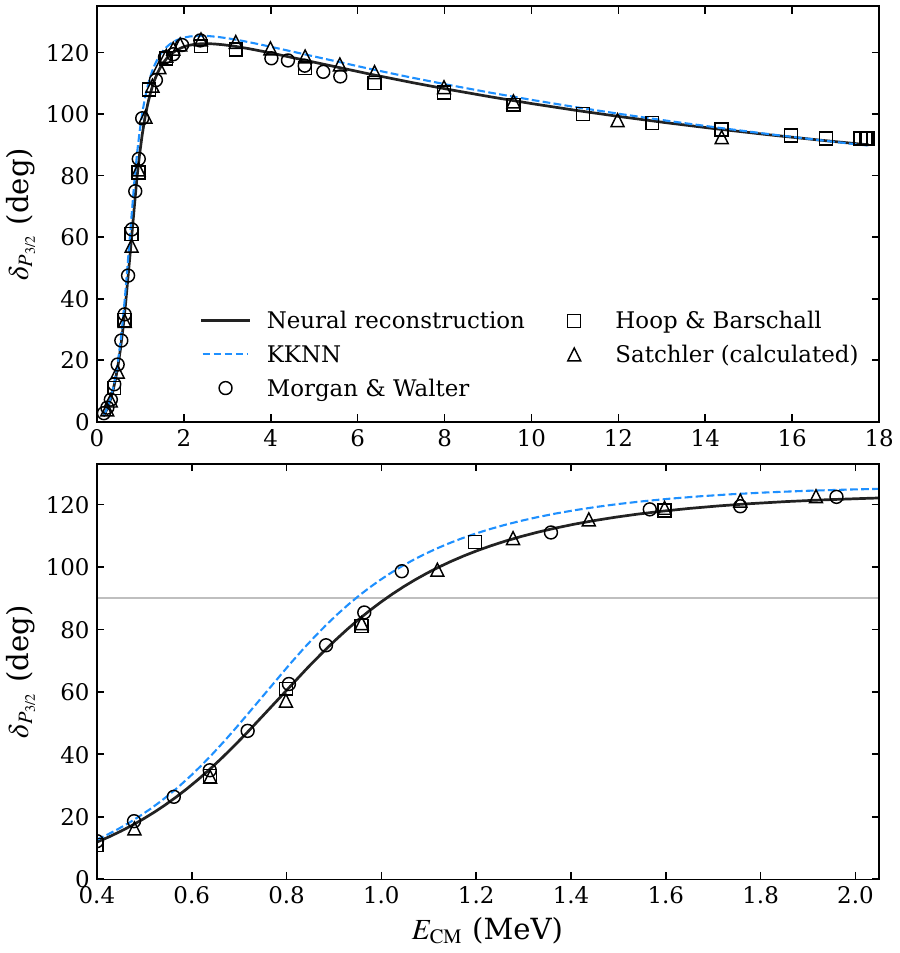}
\caption{Forward \(n+\alpha\) \(P_{3/2}\) phase shifts from the neural
reconstruction and KKNN, shown up to 20 MeV LAB. Markers are the fit
inputs: the Morgan--Walter and Hoop--Barschall measurements and the
calculated Satchler phases. The lower panel resolves the rise through
the resonance.}
\label{fig:nalpha_phases}
\end{figure}

The neural interaction follows the resonant rise and the measured
energy trend in Fig.~\ref{fig:nalpha_phases}. On the 42 measured phases, the circular RMSE is \(1.656^\circ\) for
the neural fit and \(3.857^\circ\) for the published KKNN interaction,
used here with its original parameters. The neural interaction gives
a higher cross-section peak energy and a larger FWHM than KKNN
(Table~\ref{tab:nalpha_comparison}).

\begin{figure}[H]
\centering
\includegraphics[width=\columnwidth]{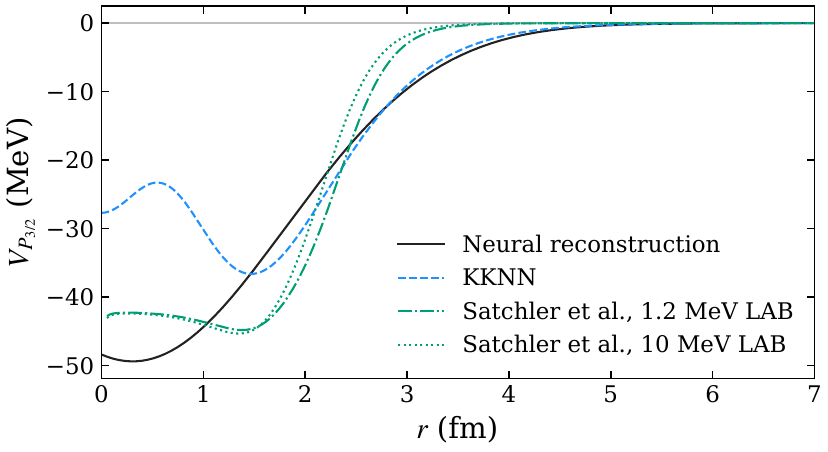}
\caption{Radial \(n+\alpha\) \(P_{3/2}\) interactions for the neural
reconstruction, KKNN, and Satchler et al. The Satchler profiles are
shown at 1.2 and 10 MeV LAB, starting at \(r=0.05\) fm because the
Thomas spin--orbit term is singular at the origin.}
\label{fig:nalpha_potentials}
\end{figure}

Figure~\ref{fig:nalpha_potentials} shows the differences in radial
shape underlying these scattering curves. The neural and KKNN
interactions are energy independent. The two Satchler profiles,
evaluated at 1.2 and 10 MeV LAB, illustrate the energy dependence
of that interaction.

\begin{table}[H]
\caption{Comparison of the neural reconstruction with the
published KKNN interaction in the $n+\alpha$
$P_{3/2}$ channel. The RMSE is evaluated against
42 empirical phase shifts from Morgan--Walter and
Hoop--Barschall. Resonance energies and widths
are given in the center-of-mass frame.}
\label{tab:nalpha_comparison}
\begin{ruledtabular}
\begin{tabular}{lccc}
Model &
\shortstack{RMSE\\(deg)} &
\shortstack{$E_{\mathrm{peak}}$\\(MeV)} &
\shortstack{FWHM\\(MeV)} \\
\hline
Neural reconstruction & 1.656 & 0.90905 & 0.88738 \\
KKNN (1979)           & 3.857 & 0.86234 & 0.78730 \\
\end{tabular}
\end{ruledtabular}
\end{table}

\subsection{Narrow resonance: \noxygen{} \dwave{}}
The \(D_{3/2}\) fit uses six phases from Fowler and Cohn~\cite{Fowler1958}
and seven from Hickey \textit{et al.}~\cite{Hickey1974}, restricted to
the first resonance (below 2.5 MeV LAB). The narrow resonance near
1 MeV LAB is resolved by tracking the continuous phase branch
through the rapid rise, with the branch controls described in
Sec.~\ref{sec:framework}.

\begin{figure}[H]
\centering
\includegraphics[width=\columnwidth]{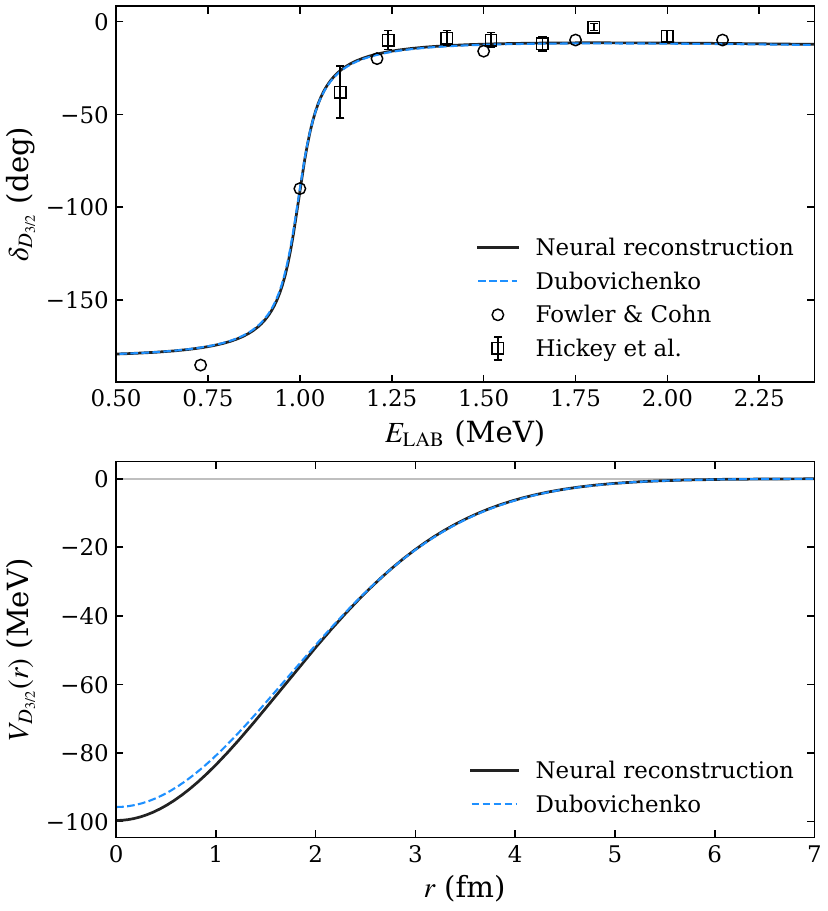}
\caption{Top: Elastic \(n+{}^{16}\mathrm{O}\) \(D_{3/2}\) phase shifts from
Fowler and Cohn~\cite{Fowler1958} and Hickey et al.~\cite{Hickey1974}, compared
with the neural reconstruction (R2) and the Dubovichenko potential~\cite{Dubovichenko2013}.
The bars give Hickey's published phase uncertainties (\(2^\circ\)--\(14^\circ\)
in this interval). Fowler and Cohn give no individual phase uncertainties.
Bottom: Corresponding effective radial nuclear potentials. The neural curve uses
the project kinetic convention, \(C_{\mathrm{neural}}=22.027964\,\mathrm{MeV\,fm^2}\),
and the Dubovichenko curve uses the published historical convention,
\(C_{\mathrm{Dub}}=21.852488\,\mathrm{MeV\,fm^2}\), where \(C=\hbar^2/(2\mu)\).}
\label{fig:oxygen_comparison}
\end{figure}

Both the representative reconstruction (R2) and the Dubovichenko Gaussian
interaction~\cite{Dubovichenko2013} reproduce this rise
(Fig.~\ref{fig:oxygen_comparison}). They give circular RMSEs of
\(5.500287^\circ\) and \(5.541805^\circ\), respectively, at the
same 13 experimental energies. With each interaction evaluated at its
own kinetic coefficient, the crossings differ by 1.1 keV LAB and the
derivative widths by 2.4 keV LAB (Table~\ref{tab:oxygen_comparison}).
The overall agreement includes a local discrepancy at 1.80 MeV LAB,
where Hickey \textit{et al.} give \(-3^\circ\pm2^\circ\). R2 and
Dubovichenko give \(-11.48^\circ\) and \(-11.80^\circ\), respectively,
on the same phase branch.

The three physically informed reconstructions (R1, R2, R3) were
initialized in a certified near-resonant region. Their circular
RMSEs are \(5.501445^\circ\), \(5.500287^\circ\), and
\(5.501783^\circ\); the corresponding \(E_{90}\) values are
1.001023, 1.001002, and 1.001024 MeV LAB, and the derivative widths
are 84.637, 85.211, and 84.534 keV LAB. The largest pairwise
circular phase difference at the 13 observation energies is
\(0.173^\circ\), and the maximum pairwise difference in the potential
on the sampled radial grid is 1.035 MeV. We take R2 as the
representative reconstruction in Fig.~\ref{fig:oxygen_comparison} and
Table~\ref{tab:oxygen_comparison}. The spread across R1--R3 measures
sensitivity to the starting point with all input phases held fixed.
Experimental uncertainties have not been propagated to \(E_{90}\)
or \(\Gamma_\delta\).

Four accepted reconstructions with a wider range of starting depths and
an adjustable envelope range (Appendix~\ref{app:ansatz}) illustrate the
radial ambiguity on a larger scale. They give \(E_{90}=0.9968\)--\(1.0010\)
MeV LAB and derivative widths of \(81.4\)--\(96.3\) keV LAB, while
\(V(0)\) ranges from \(-95\) to \(-48\) MeV.

\begin{table}[H]
\caption{Circular phase-shift RMSE at the same 13 experimental energies
and first-resonance observables for \(n+{}^{16}\mathrm{O}\) \(D_{3/2}\).
The resonance energy and derivative width
\(\Gamma_\delta=2/(d\widetilde\delta/dE_{\mathrm{lab}})\) are given in the
laboratory frame. The last row keeps the Dubovichenko potential fixed
and uses the R2 kinetic coefficient. The dash denotes an untabulated RMSE.}
\label{tab:oxygen_comparison}
\begin{ruledtabular}
\begin{tabular}{lcccc}
Model & \shortstack{\(C\)\\(MeV fm\(^2\))} & \shortstack{RMSE\\(deg)}
& \shortstack{\(E_{90}\)\\(MeV)} & \shortstack{\(\Gamma_\delta\)\\(keV)} \\
\hline
R2 & 22.027964 & 5.500 & 1.001002 & 85.211 \\
Dubovichenko & 21.852488 & 5.542 & 0.999888 & 87.625 \\
Dubovichenko & 22.027964 & --- & 1.182544 & 128.427 \\
\end{tabular}
\end{ruledtabular}
\end{table}

\section{Discussion}
In \noxygen{}, the narrow \dwave{} resonance stresses the method
differently from the broad \pwave{} benchmark: the calculated phase
is particularly sensitive to the location of the rapid rise. Agreement
with measured phases and recovery of resonance observables are separate
tests, and the in-sample RMSEs quoted here measure agreement at the
fitted energies rather than any statistical ranking of predictive
performance.

The oxygen comparison also shows why the Hamiltonian must be
specified when interpreting a narrow resonance. In the radial
Hamiltonian, \(C=\hbar^2/(2\mu)\) scales both the radial kinetic
operator and the centrifugal barrier \(C\ell(\ell+1)/r^2\).
Changing \(C\) at fixed attractive potential changes the balance
between these terms. The main comparison uses
\(C_{\mathrm{neural}}=22.027964\,\mathrm{MeV\,fm^2}\) for R2
and \(C_{\mathrm{Dub}}=21.852488\,\mathrm{MeV\,fm^2}\) for
Dubovichenko, so the agreement is between the respective
Hamiltonians (Table~\ref{tab:oxygen_comparison}).

For the fixed Dubovichenko potential, replacing the historical
convention by the project convention raises \(E_{90}\) from
0.999888 to 1.182544 MeV LAB. The \(0.80\%\) increase in \(C\) shifts the crossing by
183 keV and increases the derivative width from 87.6 to 128.4 keV LAB.
The two LAB-to-CM conversion factors agree to better than
\(2\times10^{-9}\), isolating the kinetic coefficient as the source
of the shift at the precision quoted.
Changing the depth alters the attraction relative to the kinetic
terms, and changing the range alters where that attraction competes
with the barrier. Either can move the crossing toward its original
energy for a different kinetic coefficient. This compensation at
the level of the resonance position leaves the width and radial
profile to be assessed separately. The numerical sensitivity quoted
here applies to this potential and resonance.

Radial ambiguity is familiar in inverse scattering with incomplete
phase information~\cite{ChadanSabatier1977,Alexander1996}. In this
channel the Fowler--Cohn point at 1.00 MeV LAB already locates the
\(90^\circ\) crossing. The four examples in Appendix~\ref{app:ansatz}
show how much freedom remains in the interior at comparable accuracy
on the 13 phases: changes in central depth are accompanied by changes
in radial shape, with a smaller spread at the surface. This is an
empirical ambiguity over the measured interval. Exact phase equivalence
at other energies would be a stronger condition.

The neural parametrization adjusts a flexible radial amplitude
directly through the scattering equation, with its asymptotic
decay set by the Gaussian envelope. Independent radial solvers
and convergence checks establish the numerical resolution of the
reported observables. Their physical interpretation rests on the
chosen model class: local, real, energy-independent interactions
in one partial wave per system and phase inputs over finite
energy windows. Within those assumptions, the method recovers
resonance observables and exposes the sensitivity of the inferred
radial form to the available data and Hamiltonian conventions.
Channel coupling, nonlocality, and the full spin dependence of
the nuclear interaction lie outside the information resolved
by these fits.

\section{Conclusions} 

We reconstructed a local, energy-independent effective potential for
the narrow \dwave{} resonance of \noxygen{} from elastic phase shifts,
optimizing a Gaussian-enveloped neural potential through a
differentiable variable-phase calculation and checking the result with
independent radial Schr\"odinger solvers. The broad \pwave{} resonance
of \nalpha{}, where a close neural precedent exists~\cite{Awasthi2026},
provided a benchmark against established interactions.

For \noxygen{} the representative reconstruction gives
\(E_{90}=1.001\) MeV LAB and a derivative width of \(85\) keV LAB,
close to the values from the Dubovichenko interaction evaluated with
its historical kinetic coefficient. The agreement across the three
near-resonant starts characterizes the local stability of the fit.
Across the four adjustable-range examples the central depth varies
by about \(47\) MeV despite similar crossings and phase residuals.

The method thus recovers the resonant scattering behavior of a narrow
resonance and offers a flexible, data-driven alternative to prescribing
a fixed radial form, while making explicit that the inferred interior
interaction is under-determined by the available elastic phases.

\appendix

\section{Ansatz versus converged neural solution}
\label{app:ansatz}

The reconstructed \dwave{} potential can resemble a simple Gaussian,
raising the question of whether the neural amplitude does any work or
the fit is carried by the Gaussian envelope alone. To probe this, we
examine reconstructions obtained from uninformed initializations,
rather than the near-resonant starts used for the main result, so that
the solutions are not forced toward a common interior. Four such
reconstructions (S1--S4), accepted by the same criteria as the main
result, are followed from their initial Gaussian ansatz, \(v(r)=V_c\)
with the network output set to zero, to the converged amplitude
\(v(r)=V_c+x^2 N_\theta(x)\).

\begin{figure}[H]
\centering
\includegraphics[width=\columnwidth]{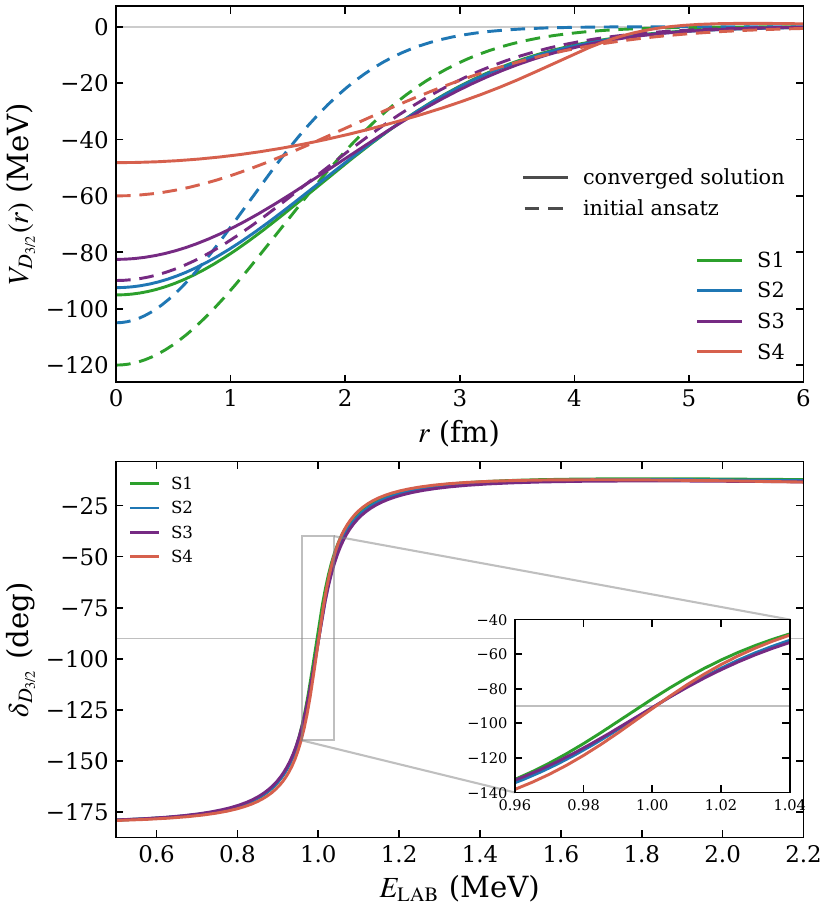}
\caption{Four accepted reconstructions of the \noxygen{} \dwave{}
resonance. Top: \(D_{3/2}\) phase shifts (no experimental points), with
a zoom on the resonant rise. Bottom: radial potentials \(V(r)\); solid
curves are the converged solutions and dashed curves their initial
Gaussian ansatz, one color per solution (S1--S4). The curves agree at
the surface and diverge in the interior.}
\label{fig:appendix_potentials}
\end{figure}

Figure~\ref{fig:appendix_potentials} shows that the four converged
potentials give comparable fits to the measured phases (upper panel) while
differing in radial shape (lower panel): they agree in the surface
region (\(r\gtrsim 2.5\) fm) but diverge in the interior, where
\(V(0)\) spans about \(-48\) to \(-95\) MeV. The converged observables
are comparable (\(E_{90}\) within \(1.00\pm0.01\) MeV LAB and circular
RMSE between \(5.6^\circ\) and \(5.8^\circ\)), so these interior differences are not resolved by the elastic
phases.

\begin{figure}[H]
\centering
\includegraphics[width=\columnwidth]{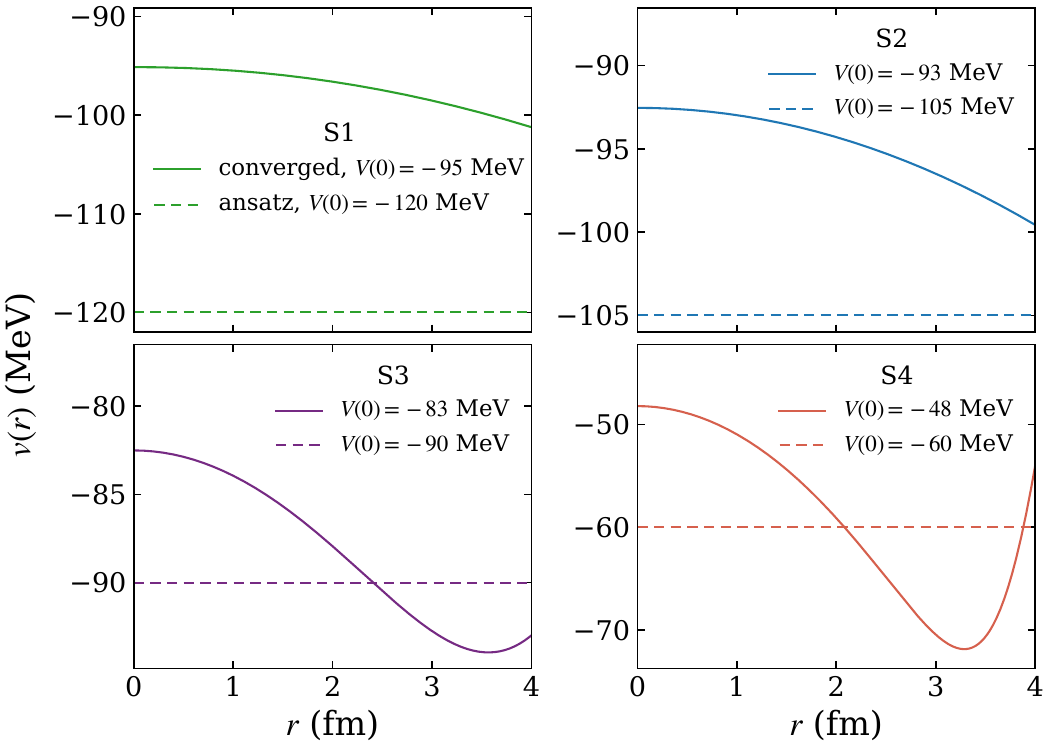}
\caption{Amplitude \(v(r)=V_c+x^2 N_\theta(x)\) for each accepted
reconstruction (S1--S4): converged (solid) versus the initial ansatz
\(v=V_c\) (dashed). The network correction is small for the deeper
starts (S1, S2) and large for the shallower start (S4).}
\label{fig:appendix_amplitude}
\end{figure}

Figure~\ref{fig:appendix_amplitude} isolates what the network learns,
comparing the converged amplitude \(v(r)\) (solid) with its ansatz
(dashed). The learned correction \(x^2 N_\theta(x)\) is not uniform: it
is small for the deeper starts (about \(2\) MeV for S1 and S2, which
therefore stay close to a Gaussian) and substantial for the shallower
start (about \(24\) MeV for S4). The network thus learns a genuine
\(r\)-dependent correction whose magnitude depends on the
initialization and compensates the initial depth to reproduce the same
resonance, consistent with the limited radial identifiability found in
the main text.

\section*{Data and code availability}
The phase-shift tables, source code, and machine-readable data supporting
the figures and numerical results are available from the corresponding
author upon reasonable request.

\begin{acknowledgments}
EO thanks the Funda\c{c}\~ao Cearense de Apoio ao Desenvolvimento Cient\'ifico e Tecnol\'ogico (FUNCAP), through grant BP6-0241-00335.01.00/25.
\end{acknowledgments}

\bibliographystyle{apsrev4-2}
\bibliography{references}
\end{document}